\documentclass[aps,prl,reprint]{revtex4-1}

\usepackage{amsmath}
\usepackage{amssymb}
\usepackage{amstext}
\usepackage{bm}
\usepackage{graphicx}
\usepackage{array}
\usepackage{multirow}
\usepackage{lipsum}
\usepackage{color}
\usepackage{ulem}
\usepackage{mathtools}

\usepackage{subfig}
\usepackage{graphicx}

\begin{document}
\title{Quantitative orbital angular momentum measurement of perfect vortex beams}
\author{Jonathan Pinnell}
\author{Valeria Rodriguez-Fajardo}
\author{Andrew Forbes}
\email{Corresponding author: andrew.forbes@wits.ac.za}
\affiliation{School of Physics, University of the Witwatersrand, Johannesburg 2000, South Africa}




\begin{abstract}
Perfect (optical) vortices (PVs) have the mooted ability to encode orbital angular momentum (OAM) onto the field within a well-defined annular ring.  Although this makes the near-field radial profile independent of OAM, the far-field radial profile nevertheless scales with OAM, forming a Bessel structure.  A consequence of this is that quantitative measurement of PVs by modal decomposition has been thought to be unviable.  Here, we show that the OAM content of a PV can be measured quantitatively, including superpositions of OAM within one perfect vortex. We outline the theory and confirm it by experiment with holograms written to spatial light modulators, highlighting the care required for accurate decomposition of the OAM content.  Our work will be of interest to the large community who seek to use such structured light fields in various applications, including optical trapping and tweezing, and optical communication.
\end{abstract}


\maketitle

Complex structured light fields have been demonstrated in a myriad of forms \cite{Forbes2016} and found many applications covering both the classical and quantum regimes \cite{roadmap}.  Foremost amongst these are helical phase structures that carry orbital angular momentum (OAM) \cite{Allen1992,OAM1}.  Such modes have an azimuthal phase structure of $\exp(i\ell \phi)$ where $\phi$ is the azimuthal angle and carry $\ell \hbar$ of OAM per photon, with $\ell$ an integer.  Such OAM modes are conventionally created by passing a Gaussian beam through a helical phase, often on a spatial light modulator (SLM) \cite{White}, to create a vortex beam as an approximation to the azimuthal modes in the Laguerre-Gaussian basis.  This approach produces hypergeometric modes \cite{karimi2007hypergeometric} with little power in the desired mode of helicity $\ell$ and radial order $p = 0$ \cite{sephton2016revealing}.  Further, the ring of light about the phase singularity has a radius of $r_\ell = w_0 \sqrt{|\ell|/2}$, where $w_0$ is the waist radius of the embedded Gaussian beam, resulting in a ring of light that scales in size with OAM (the second moment size likewise scales as $w_\ell = w_0 \sqrt{|\ell| + 1}$).  To overcome this the concept of a perfect (optical) vortex (PV) was introduced \cite{ostrovsky2013generation}.  The PV is an annular ring of fixed radius and thickness with an encoded helicity.  Such annular structures are well-known as the Fourier transform of Bessel beams \cite{Durnin1987,vaity2015perfect}, so the price to pay for an OAM independent radial scale in the near-field is an OAM dependent radial scale in the far-field \cite{Vasilyeu2009A}.  

PVs with OAM have been extensively created \cite{garcia2014simple,chen2015creating,chen2015generation,banerji2016generating,li2016generation,liu2017generation,tkachenko2017possible,zhang2018generating} and applied \cite{chaitanya2016efficient,fu2016perfect,jabir2016generation,zhang2016perfect,li2018controllable,shao2018free}, but measurement techniques are still in their infancy \cite{ma2017situ}.  In contrast, OAM has been measured \textit{qualitatively} using mode sorters for Laguerre-Gaussian and helical beams \cite{Berkhout2010A,LG,ruffato2016diffractive,ruffato2018compact}, extended to radial modes \cite{zhou2017sorting,gu2018gouy}, with adapted approaches to find the mode indices for Bessel-Gaussian \cite{dudley2013efficient,dudley2013generating,trichili2014detection} and Hermite-Gaussian modes \cite{HG}.  This detects the mode but cannot return the full information of the field.  Modal decomposition on the other hand is a generic approach to reconstruct any optical field \textit{quantitatively} \cite{Duparre2000A,litvin2012azimuthal,Flamm2012B,Flamm2013A}.  It has been exploited for the measurement of phase and wavefronts \cite{Schulze2012A}, OAM density \cite{Schulze2013} and beam quality factors \cite{Schmidt2011real,Flamm2012B}.  

Here, we demonstrate that the OAM content of a PV can be measured quantitatively and outline the theory and experiment to do so.  We show accurate reconstruction of the OAM properties and show that this approach works even for superpositions of OAM PVs.  In the process we highlight the prior misconceptions that prohibited this analysis, namely, that the modal expansion is required to be into an orthogonal or orthonormal basis, of which PVs and Bessel beams are not.  We show how to overcome this, paving the way for the detection and characterization of such structured light fields.

\begin{figure}[t]
	\centering
	\includegraphics[width=\linewidth]{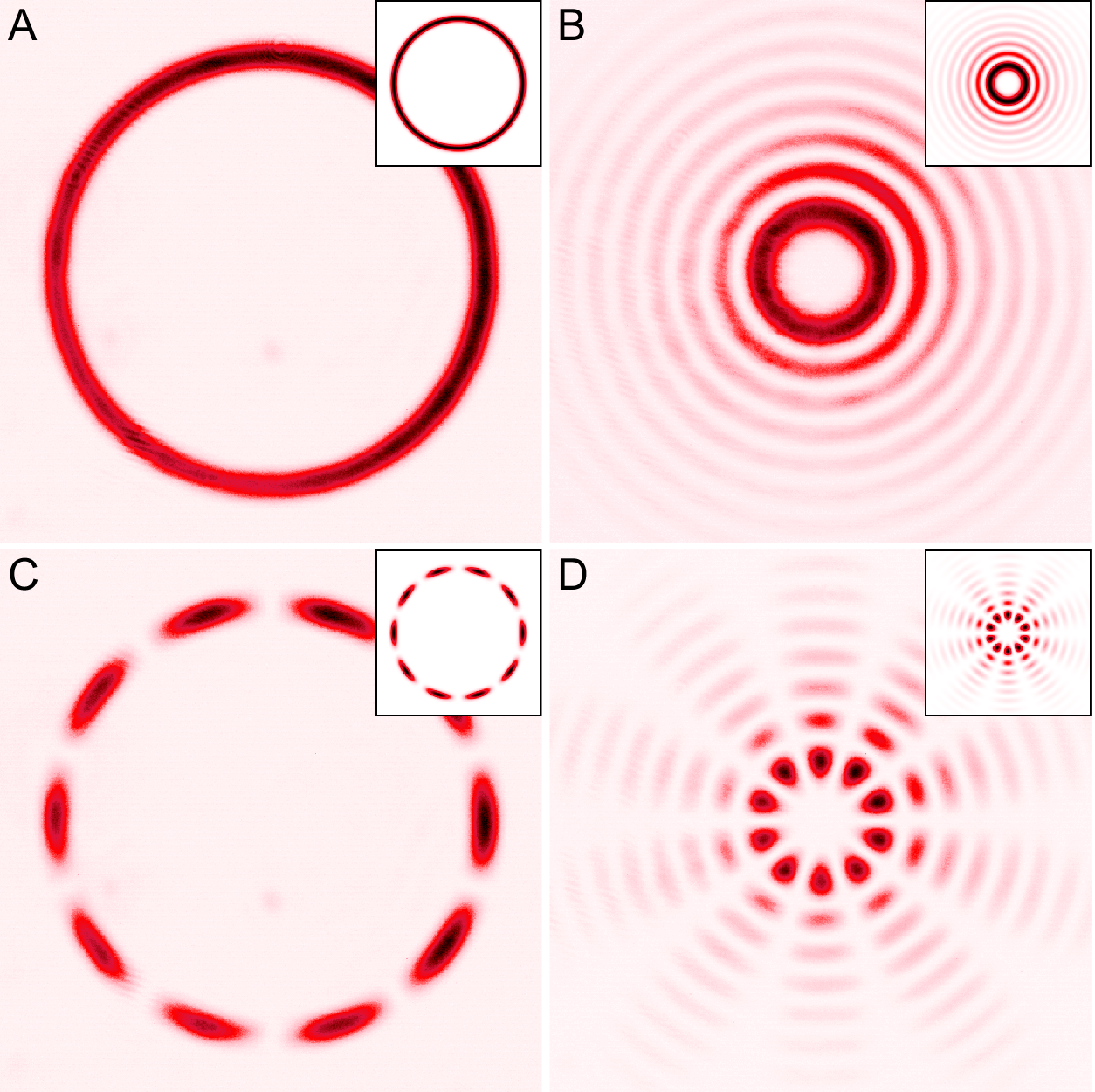}
	\caption{Generation of PV beams with $k_r = 15\, \text{mm}^{-1}$, $w_0 = 2\,\text{mm}$ and $f = 1000\,\text{mm}$. Transverse intensities for $\ell=5$ are shown in A,B and a superposition of $\ell=5$ and $\ell=-5$ is shown in C,D. Images A,C correspond to NF transverse intensities and B,D to FF intensities. Insets display simulated images.}
	\label{fig:PVs}
\end{figure}

We begin with the generation of PV beams. As stated earlier, PVs are simply the Fourier transform of the well-studied Bessel beams. Here, we consider a PV to be the near-field (NF) spatial profile and hence the far-field (FF) profile is the corresponding Bessel beam. An ideal PV resulting from an ideal Bessel beam would be an annular ring of infinitesimal thickness and radius $r_r = k_r f/k$, where $k_r$ is the radial wave number of the Bessel beam, $f$ is the focal length of the Fourier lens and $k$ is the wave number of the light. An ideal Bessel beam cannot be experimentally realised and so we turn to it's finite-energy approximation: the Bessel-Gauss beam whose transverse electric field in cylindrical coordinates $(r,\phi)$ at the waist plane is described as,
\begin{equation} \label{eq:BG}
    E(r,\phi) \propto J_\ell(k_r r)\, \exp\left( -\frac{r^2}{w_0^2} \right) \, \exp(i\ell\phi)  \,,
\end{equation}
where $J_\ell(\cdot)$ is the Bessel function of the first kind of order $\ell$. The PV from a Bessel-Gauss beam is no longer infinitesimally thin, but instead has thickness $r_t = 2 f / k w_0$. The electric field of this experimentally realisable PV at the waist plane is given by \cite{vaity2015perfect},
\begin{equation} \label{eq:PV}
E(r,\phi) \propto \, \frac{w_0}{r_t} \, \exp\left( -\frac{r^2 + r_r^2}{r_t^2} \right) \, I_\ell \left( \frac{2 r_r r}{r_t^2} \right) \, \exp(i\ell\phi)  \,,
\end{equation}
where $I_\ell(\cdot)$ is the modified Bessel function. In Fig.~\ref{fig:PVs}, the theoretical and experimental transverse intensities at the waist plane are shown for the NF and FF region for single OAM modes and superpositions of OAM, the NF corresponding to the PV and the FF to the Bessel-Gaussian beam. The correlation coefficient between the  theoretical and experimental intensities is between $97-99\%$ in all cases, indicating that the experimental beams were generated with high fidelity. 

We now turn to the quantitative OAM detection of PVs in both the NF and FF regions. There is a misconception that the OAM content of PVs cannot be decomposed in the same way as regular vortex beams (namely with a spiral phase plate and a lens). The reasoning is that a spiral phase-plate by itself cannot transform the PV back into a Gaussian; In the NF additional transformation optics such as an exicon would be needed to ``collapse'' the ring, while in the FF the Bessel beams do not form an orthonormal basis. Although true, this reasoning misses the full picture of the modal decomposition method. We'll now show that the OAM of a PV can indeed be detected quantitatively. Further, we'll show that, in principle, the requirement for quantitative OAM detection of any beam is that its azimuthal phase is variable separable from its amplitude.

In the general case, suppose that one wishes to express some initially unknown field $E(\mathbf{x})$ in some basis $\Phi_n(\mathbf{x})$, where the components of $\mathbf{x}$ are transverse spatial coordinates. The task is then to find the expansion coefficients ($c_n$) in
\begin{equation} \label{eq:exp}
E(\mathbf{x}) = \sum c_n \, \Phi_n(\mathbf{x}) \,.
\end{equation}
\begin{figure}[tb]
	\centering \includegraphics[width=\linewidth]{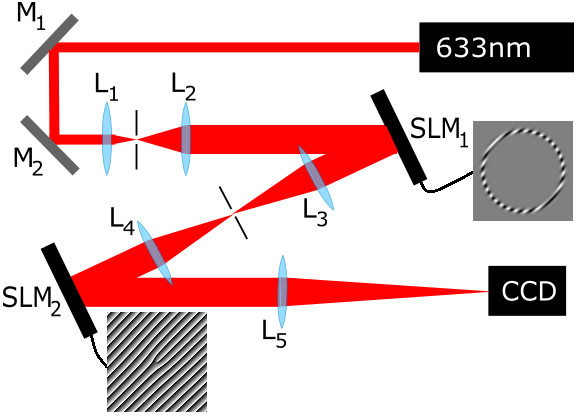} 
	\caption{Schematic of the  experimental set-up; $M_i$ denote mirrors, $L_i$ are lenses and SLM$_i$ are spatial light modulators.} 
	\label{fig:Expsetup}
\end{figure}
Using the orthogonality of the basis functions, the expansion coefficients can be computed from
\begin{equation}
c_n = \int d^2 \mathbf{x} \, E(\mathbf{x}) \, \Phi^*_n(\mathbf{x}) \,.
\end{equation}
How is this done optically? First, let the field interact with an optical element whose transmission function is $T(\mathbf{x}) = \Phi^*_n(\mathbf{x})$ (such as a SLM displaying a computer generated hologram). The field immediately after this element is then $E(\mathbf{x})\Phi^*_n(\mathbf{x})$. Now consider what happens as this modified field passes through a Fourier lens. At the back focal plane of the lens, we have
\begin{align}
E(\mathbf{x}) \xrightarrow[]{\text{SLM+lens}} \mathcal{E}(\mathbf{k}) \propto \int d^2 \mathbf{x} \, E(\mathbf{x})\, \Phi^*_n(\mathbf{x}) \exp \left(-i \frac{k_0}{f} \, \mathbf{k} \cdot \mathbf{x} \right) \,,
\end{align} 
where $k_0 = 2\pi/\lambda$ is the initial field's wave number and $f$ is the focal length of the lens. If we restrict our gaze to the on-axis intensity of the beam (with a camera) where $\mathbf{k} = 0$ then
\begin{align} \label{eq:overlapInt}
|\mathcal{E}(\mathbf{0})|^2 &\propto \left| \int d^2\mathbf{x} \,E(\mathbf{x}) \, \Phi^*_n(\mathbf{x}) \right|^2 =  |c_n|^2 \,.
\end{align} 
This, then, is an effective way of finding the magnitude of the expansion coefficients in Eq.~\ref{eq:exp} optically. 

Returning to the specific case of PVs, begin by factorising out the azimuthal phase in the field amplitude (Eq. \ref{eq:BG} and Eq. \ref{eq:PV}), writing $E(r,\phi) = R_\ell(r) \exp(i\ell\phi)$ and use as a basis the OAM eigenstates $\Phi_n(\phi) = \exp(i n \phi)$. It then follows that
\begin{align}
|\mathcal{E}(\mathbf{0})|^2 &\propto \left| \int_0^\infty r \, dr\, R_\ell(r) \int_0^{2\pi} d\phi\, \exp\left[i (\ell-n)\phi\right] \right|^2 \label{eq:PVoverlap} \,,\\
& = | \gamma_\ell |^2\,\, \delta_{\ell,n} \label{eq:delta}  \,,
\end{align} 
where the last relation holds because the radial integral of the PV amplitude evaluates to a constant $\gamma_\ell$ (which may depend on $\ell$ in general). This result shows that there will be an on-axis intensity in the Fourier plane if and only if the helical phase of the PV has been unwrapped correctly (provided that the radial integral is non-zero). Further, this decomposition should hold for any physically realisable beam whose azimuthal phase can be factorised from the amplitude; PVs are just a special case. Notice that if the initial field is a superposition of PVs with different $\ell$ values then Eq.~\ref{eq:delta} becomes a sum of such terms and the decomposition is still effective.

\begin{figure}[tb]
	\centering
	\includegraphics[width=\linewidth]{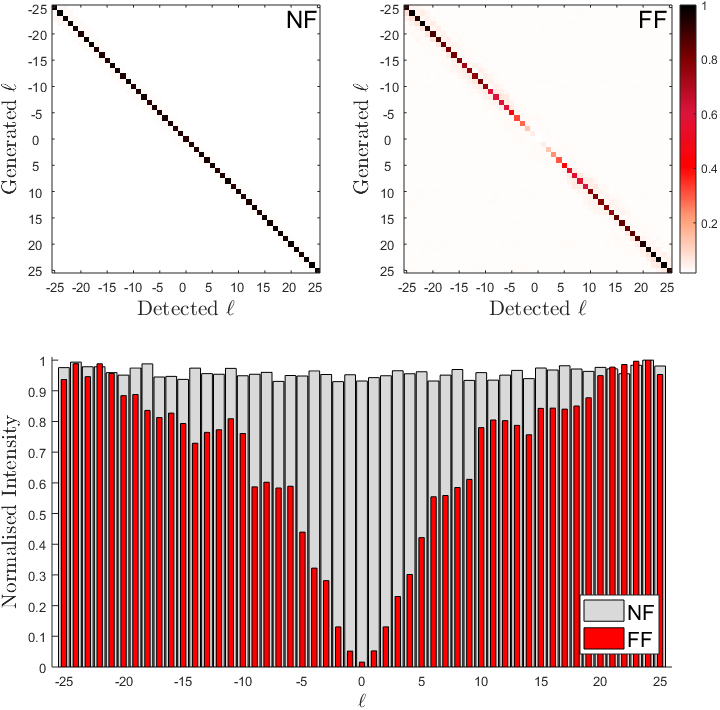}
	\caption{Experimental results of the single mode OAM decomposition of PVs. The first two images show the cross-talk matrices in the NF and FF. The plot beneath shows the diagonal part of these matrices, highlighting the variation of the on-axis intensity.}
	\label{fig:CrossTalks}
\end{figure}

To confirm the OAM decomposition, we perform the experiment shown schematically in Fig.~\ref{fig:Expsetup}. The beam from a He-Ne laser is expanded and collimated onto the first SLM which is then relayed to the second SLM using a $4f$ lens system. A Point Grey Firefly camera is placed at the Fourier plane of the last lens for the detection of the on-axis intensity. The two phase-only Holoeye Pluto SLMs carry the bulk of the workload: one generates the PV directly using complex-amplitude modulation \cite{arrizon2007cam} and the other scans through a set of forked holograms (one at a time) encoding $\Phi_n^*$. If desired, one can multiplex many holograms onto the detection SLM so as to perform the OAM measurement in a single-shot.

In the first row of Fig.~\ref{fig:CrossTalks}, the results of the single mode decomposition are summarised in a cross-talk matrix to quantify the effectiveness of the detection system. Ideally, one would obtain an identity matrix indicating that the detection system can successfully decompose what was generated. The beam parameters of all experimentally generated PVs were kept constant and only the topological charge $\ell$ was varied. In addition, all generated beams were normalised to unit amplitude. We find excellent agreement with Eq.~\ref{eq:delta} over a large range of OAM values: $\ell\in[-25,25]$. This range was chosen since the beam size of the PV in the FF for $|\ell| > 25$ was larger than the numerical aperture of the system. We can conclude that the OAM mode cross-talk in the detection system is minimal for both the NF and FF regions, owing to the negligible off-diagonal matrix components. Note how the on-axis signal diminishes for decreasing $|\ell|$ when decomposing in the FF (where the vortex ceases to be ``perfect"), as can be seen in the second row of Fig \ref{fig:CrossTalks}. For $|\ell|=0$, the signal is essentially at background level. 

\begin{figure}[tb]
	\centering
	\includegraphics[width=\linewidth]{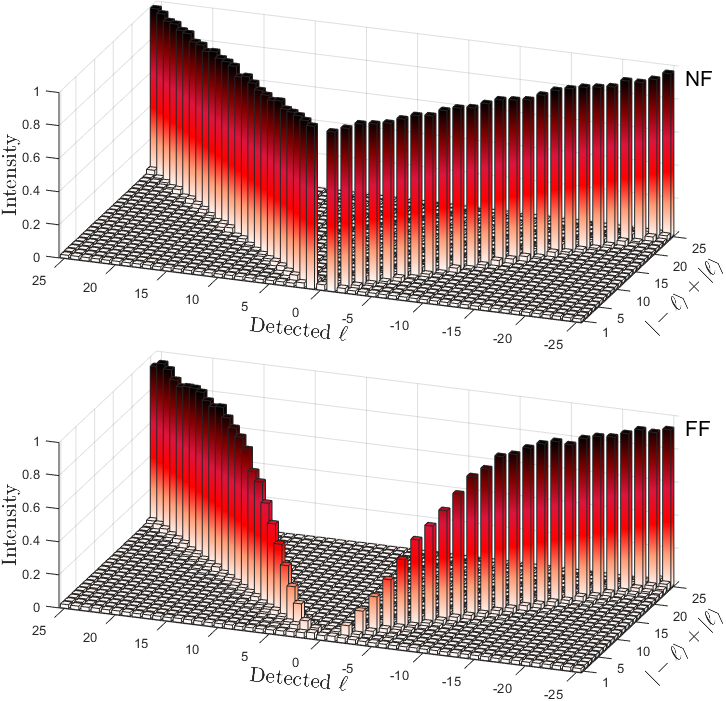}
	\caption{Experimental decompositions of symmetric superpositions of PVs in the NF and FF.}
	\label{fig:Super}
\end{figure}

Figure \ref{fig:Super} shows the decomposition of various symmetric superpositons of PVs in the NF and FF regions, where each row corresponds to a generated OAM mode of the form $|-\ell\rangle + |\ell\rangle$. As claimed earlier, the decomposition is still effective as there is no ambiguity as to which OAM modes contributed to the OAM spectrum; this is especially true in the NF where the PV exists. For similar reasons as before, when decomposing in the FF the on-axis signal falls to background level as $|\ell|$ decreases.

\begin{figure}[tb]
	\centering
	\includegraphics[width=\linewidth]{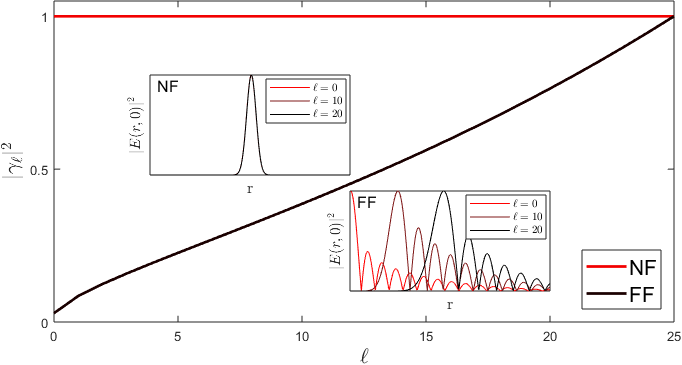}
	\caption{Variation of the (normalised) on-axis intensity $|\gamma_\ell|^2$ with $\ell$. Insets show the radial intensity profile of the PV in the NF and FF regions: the former is OAM independent while the latter is not. }
	\label{fig:gamma}
\end{figure}

We can explain the signal drop off with $\ell$ in the FF (where the PV does not exist) by returning to the field equations for the NF and FF beams.  The theoretical variation of the on-axis intensity with the mode being decomposed is highlighted in Fig. \ref{fig:gamma} where $|\gamma_\ell|^2$ in Eq. \ref{eq:delta} is computed explicitly for the OAM values used in the experiment. This signal variation is the cost for having an OAM dependent radial scale, as the insets of Fig. \ref{fig:gamma} show, again highlighting the fact that the PV exists only in the NF. Since cameras have finite dynamical range, this fact makes the OAM decomposition of PVs in the FF difficult when utilising a large range of $|\ell|$ values.  In general, when decomposing into a non-orthonormal basis (such as the OAM eigenstates which are not a basis of the transverse plane), one has to apply appropriate correction factors to the raw OAM spectrum to get the true spectrum. This is due to the previously mentioned problem of varying on-axis signal between different decomposed modes. In a physical sense, energy normalisation needs to be imposed for each mode. In a mathematical sense, the constant of proportionality in the overlap integral in Eq. \ref{eq:overlapInt} depends on the mode parameters being decomposed ($\ell$ in our case) and will likely be different across the spectrum. This needs to be corrected for in order to obtain the true spectrum. This is not such a problem for symmetric OAM superposition since the correction factors would scale the spectrum evenly ($\gamma_\ell = \gamma_{-\ell}$). For asymmetric superpositions, however, correction factors would need to be applied with care, calculated by evaluating the appropriate overlap integral of the system (i.e. computing $\gamma_\ell$). In principle, these factors can be obtained experimentally from the inverse of the diagonal part of the cross-talk matrix. This is because each incoming mode is engineered to have one expansion coefficient of unit magnitude and so the correction factor can be easily determined from the inverse of the measured coefficient.
For the case of PVs, we see from our results that the on-axis signal changes negligibly with $\ell$. This is due to the fact that the beam width (and thus the energy since the radius is constant) of a PV changes negligibly with $\ell$.  This is shown in our NF results.  This showcases a clear advantage for using PVs as ``carriers" of OAM over any other set of beams. When Fourier transformed to the FF, creating Bessel beams, the advantages are lost. 

In conclusion, we have examined the theory and shown experimental results which demonstrate the detection of OAM in PVs with a simple modal decomposition technique. In all cases, the decomposition of PVs in the NF posed fewer problems and yielded more reliable results than FF PVs: no energy normalisation or OAM spectrum correction factors need be imposed post hoc. Further, the scaling of NF PVs with $\ell$ is minimal, unlike regular vortex beams whose radius scales as $\sqrt{|\ell|}$. This vastly increases the range of usable OAM modes and would be useful in any application where regular vortex modes are currently being used, such as optical trapping, tweezing and communications. 

\bibliography{Reference_File}




\end{document}